\documentclass[runningheads]{llncs}
\usepackage[T1]{fontenc}
\usepackage{graphicx}
\usepackage[
backend=biber,
style=numeric, maxbibnames=20,
sorting=ynt
]{biblatex}
\usepackage{color}

\begin{document}

\title{Ning Cai: A Tribute to a Pioneering Scholar\protect\\ in Information Theory}

\author{Ingo Althöfer \and Holger Boche \and Christian Deppe \and \protect\\ Ulrich Tamm \and Andreas Winter \and Raymond W. Yeung}

\authorrunning{I. Althöfer, H. Boche, C. Deppe, U. Tamm, A. Winter, R.W. Yeung} 

\institute{}

\maketitle

\renewcommand{\abstractname}{}

\begin{abstract}
It is with heavy hearts that we mourn the passing of Ning Cai, a luminary whose pioneering spirit illuminated the realms of network coding and beyond. On May 25, 2023, at the age of 75, Prof. Cai bid farewell, leaving behind a profound legacy that continues to resonate across generations of researchers.
His contributions spanned a vast spectrum, from the groundbreaking explorations in network coding to the intricate realms of quantum information theory. Ning's indelible mark on the academic landscape is a testament to his unwavering dedication and relentless pursuit of knowledge. Among his many accolades, Ning's seminal works garnered widespread recognition, exemplified by the prestigious 2005 IEEE Information Theory Society Paper Award for his work "Linear Network Coding." Furthermore, his enduring impact was underscored by the 2018 ACM SIGMOBILE Test-of-Time Paper Award, bestowed upon his paper "Network Information Flow." In addition to his scholarly achievements, Ning's unwavering commitment to mentorship has left an indelible mark on countless aspiring scholars. His guidance and wisdom continue to inspire and guide future generations in their scholarly pursuits. As we bid farewell to a titan in the field, let us cherish the legacy of Ning Cai, whose brilliance and generosity of spirit will forever endure in the annals of academia.
\end{abstract}

Ning Cai hailed from an academic lineage, both of his parents serving in positions at university. His father, Cai Changnian, a professor who later ascended to the role of Vice President of Beijing Institute (now University) of Posts and Telecommunications, notably penned the first Chinese textbook on information theory. Meanwhile, his elder sister, Cai Anni, pursued an academic career as professor at the same institution.

Ning pursued his passion for mathematics, earning his Bachelor's degree in 1982 from Beijing Normal College, followed by a Master's degree in 1984 from the Chinese Academy of Sciences, also located in Beijing. His Master thesis advisor was Zhang Zhaozhi, a top expert in cryptography in China. After obtaining his Master's degree, he became an assistant researcher at the Chinese Academy of Sciences.

In 1986, a pivotal shift occurred in both Ning's personal and academic trajectory. Rudolf Ahlswede's visit to China for a lecture series marked the beginning of a significant chapter in Ning's life. During Ahlswede's visit, he posed several challenging problems, one of which Ning Cai successfully tackled, igniting a promising collaboration. This prompted Ahlswede to extend an invitation for Ning to join him at the University of Bielefeld to pursue his Ph.D. In 1987, Ning relocated from China to Germany, establishing Bielefeld as his new home until his passing. Over time, he even acquired German citizenship.

Upon Ning's arrival, significant transitions were unfolding within Ahlswede's academic circle at Bielefeld. The initial cohort of Ahlswede's associates had recently moved on to new endeavors: Gunter Dueck embarked on a prosperous managerial career at IBM Germany, while Ingo Wegener and others secured professorial roles in Computer Science departments. In this evolving landscape, Ingo Althöfer assumed the role of assistant professor, and Zhen Zhang, fresh from completing his Ph.D. under Toby Berger at Cornell, was appointed as the second assistant under Ahlswede's tutelage.

Upon his return from China, Ahlswede brought back three Ph.D. students: Ning Cai, Ye Jiang Ping, and Fangwei Fu, all of whom joined Bielefeld's mathematics department in 1987. Ulrich Tamm, in the midst of completing his diploma thesis, rounded out the research group. A year later, Klaus-Uwe Koschnick rejoined as a Ph.D. student.

Observers noted Ning's seniority compared to Ye and Fangwei, later discovering the somber reason behind it. Ning commenced his university studies at the age of 30 due to past hardships he was forced to endure during China's post-Cultural Revolution era. This included forced labor in rural areas, an experience Ning likened to a concentration camp.

While Ye Jiang Ping and Fangwei Fu returned to China following the completion of their Ph.D. studies in 1988, Ning chose to remain in Bielefeld. In 1989, the mathematics department saw the establishment of the special research unit (SFB) "Discrete Structures in Mathematics," overseen by Rudolf Ahlswede, who led two research groups focusing on information theory and combinatorics, respectively. One of these positions was designated for Ning Cai. During the latter part of 1988, Ning worked with Vera Pless in Chicago, where they collaborated on a joint paper with Brualdi.

In his Ph.D. thesis titled "4-Words and Diametrical Inequalities," Ning delved into communication complexity. Rudolf Ahlswede's keen interest lay in the Hamming distance, whose communication complexity could be lower bounded through monochromatic rectangles, denoted as code pairs $A \times B$, where a function maintains a constant value. After grappling with increasingly intricate induction proofs involving larger alphabet sizes, Ahlswede tasked his students with devising a single-letter characterization to streamline the analysis from a word of $n$ letters to a single letter. Ning eventually achieved this feat by relaxing the requirement from a "constant value" to the "4-word-property" ($f(a,b) - f(a',b)-f(a,b')+f(a',b') = 0$ for any $a, a' \in A$ and $b,b' \in B$). This breakthrough enabled the determination of the communication complexity of the Hamming distance and other functions asymptotically - and precisely, if only one of the two communicators needed to be informed about the result.

\begin{figure}
    \centering
    \includegraphics[width=10cm]{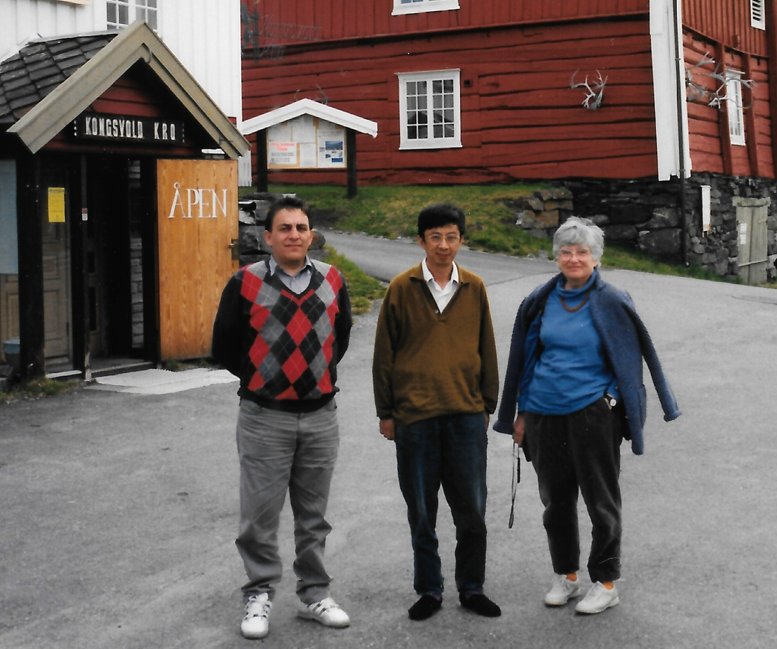}
    \caption{Levon Khachatrian, Ning Cai, Vera Pless (from left to right) in Norway 1994.}
    \label{fig:Norway-1994}
\end{figure}

Communication complexity was introduced and first considered by Andrew Yao. However, in Yao's original model, both communicators must ultimately be aware of the result. Upon returning from Chicago, Ning embarked on developing a single-letter characterization for this scenario as well. Once again, he succeeded through a relaxation approach.  

By refraining from summing up the individual components' results and instead focusing solely on the "vector" $f^n$ of the componentwise evaluation of a fundamental function $f$, the communication complexity of this basic function $C(f)$ can be compared to the average complexity per component of the so-called vector-valued function.
Ning also provided the example of the componentwise minimum of binary vectors, where this average complexity is 
$$\overline{C}(f) = \lim_{n \rightarrow \infty} \frac1n C(f^n) = \log_2 3,$$ 
which notably falls below the communication complexity $C(f)=2$ for the fundamental function $f$.

With his research Ning was far ahead of his time. The quantity, $\lim_{n \rightarrow \infty} \frac{1}{n} C(f^n)$, later assumed significant importance in Computer Science, where it is referred to as the amortized communication complexity and plays a central role in Avi Wigderson's famous direct-sum conjecture.

Ning's lower bound technique using prefix codes is also the first application of protocol compression later studied by Braverman et al. for probabilistic communication protocols.

Ning's work on communication complexity is surveyed more detailed in Ulrich Tamm's contribution in this volume. 

Between 1989 and 1997, Ning Cai, alongside Rudolf Ahlswede and Zhen Zhang, delved into various challenges applying combinatorial methods to information theory. Their investigations encompassed diametrical inequalities, posets and lattices, memories, among other things.

Of particular significance in information theory are the collaborative papers by Ning and Rudolf Ahlswede on the Arbitrarily Varying Channel (AVC), wherein they successfully tackled numerous complex problems.

A pivotal condition emerged regarding the capacity region of an arbitrarily varying multiple access channel (AVMAC) having a non-empty interior. This condition gained prominence after Jahn's 1981 determination of the average error capacity region of AVMAC under the stipulation of a non-empty interior for the capacity region. Nine years later, Gubner's 1990 findings provided a necessary condition for an AVMAC to possess a non-empty interior, accompanied by the conjecture that it would suffice. Despite ongoing challenges in decoder design and decoding error estimation, Ahlswede and Cai finally proved the conjecture in 1999, after nine more years. Furthermore, in the same work's second part, they demonstrated that enlarging a capacity region with an empty interior to the random code capacity region could significantly enhance transmission through AVMAC, when accessing a correlated source via two encoders.

The quest to determine the maximum error capacity of AVC with feedback posed a formidable challenge. After Ahlswede's partial solution for a special case in 1973, progress stagnated, leaving the problem unresolved for 27 years until Ahlswede and Cai accomplished a comprehensive solution in 2000.

In 1991, Ning resolved a conjecture on list decoding for AVC posited by Pinsker in 1990. Additionally, he unveiled that a correlated source accessed by the sender and receiver could elevate AVC capacity from zero to the random code capacity (Ahlswede/Cai, 1997).

During the early 1990s, Zhen Zhang and Ingo Althöfer secured professorial positions in Los Angeles and Jena, respectively. This transition left Rudolf Ahlswede, Ning Cai, and Ulrich Tamm, to be joined by Levon Khachatrian, Bernhard Balkenhol, and later Vladimir Balakirski and Harout Aydinian. Moreover, Kingo Kobayashi spent two years contributing to their endeavors in Bielefeld.

Beyond the confines of the special research unit, collaborative projects with Han Vinck in Essen flourished. This collaboration facilitated visits to Japan and attendance at conferences in Essen, Eindhoven, Moelle, and the ISITs in Trondheim and Ulm, documented through photographs taken during these trips.

\begin{figure}
    \centering
    \includegraphics[width=10cm]{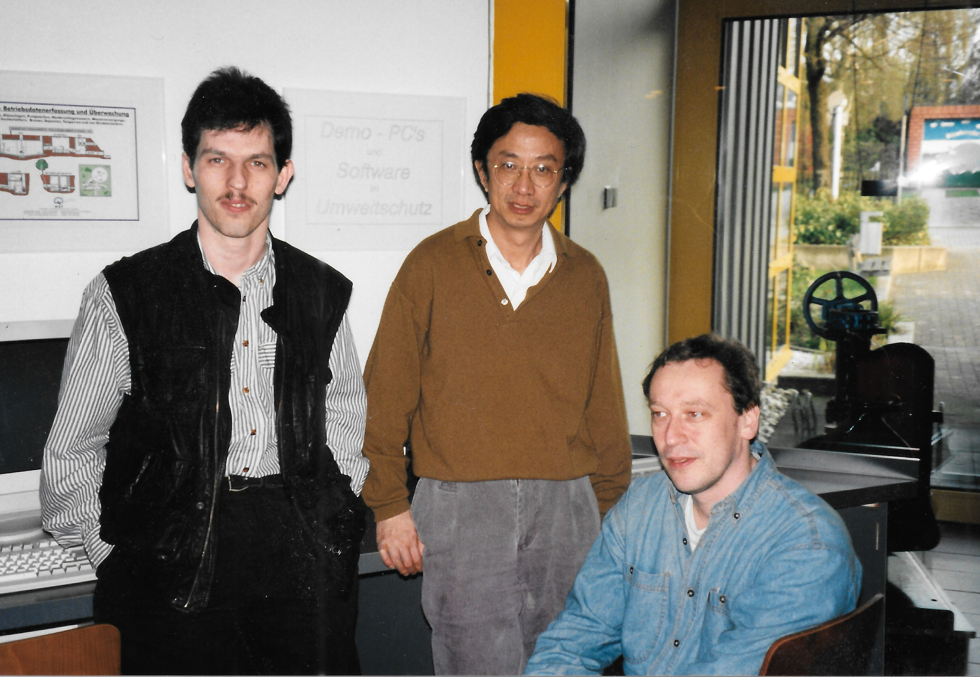}
    \caption{Bernhard Balkenhol, Ning Cai, Ulrich Tamm (from left to right) visiting Han Vinck in Essen 1997.}
    \label{fig:Essen-1997}
\end{figure}

The SFB years were marked by excitement, particularly with the recent opening of borders in Europe, which facilitated the influx of guests, notably from Russia and Hungary. Rudolf Ahlswede's robust connections to the Institute for Problems of Information Transmission in Moscow and the Renyi Institute in Budapest required endless visa acquisitions for research trips. Notably, esteemed figures like Paul Erdös and Mark Pinsker were frequent visitors to Bielefeld during this vibrant period.

Rudolf Ahlswede maintained an ambitious vision that set him apart from many of his colleagues in Bielefeld, ensuring that his group remain consistently engaged. Responsibilities were clearly delineated within the team: Ning focused on information theory research, Levon on combinatorics, Bernhard managed IT infrastructure and programming tasks, while Ulrich handled teaching, guest relations, administration, and the organization of seminars and workshops.

The team's output, both in terms of quality and quantity, was notably impressive. Ning's diligent efforts yielded fine and original results, exemplified by his work on communication complexity and arbitrarily varying channels (AVC).

Ning's research reached new heights with network coding, a field he pioneered alongside Ahlswede, Li, and Yeung. In 1997, Ahlswede invited Raymond Yeung for a two-week visit to the Bielefeld University as part of the Collaborative Research Center (SFB) before the ISIT in Ulm, Germany. During this period, Ning and Raymond Yeung embarked on exploring the max-flow min-cut theorem for single-source network coding. Subsequently, Bob Li from Chinese University of Hong Kong joined the endeavor, contributing the concept of the "butterfly network." Their seminal paper "Network Information Flow," published in the IT Transactions in 2000, garnered the 2018 ACM SIGMOBILE Test-of-Time Paper Award and has amassed over 10,000 citations on Google Scholar. For his contributions to network coding, Ning received the prestigious 2016 IEEE Eric E. Sumner Award, shared with Bob Li and Raymond Yeung.

\begin{figure}
    \centering
    \includegraphics[width=10cm]{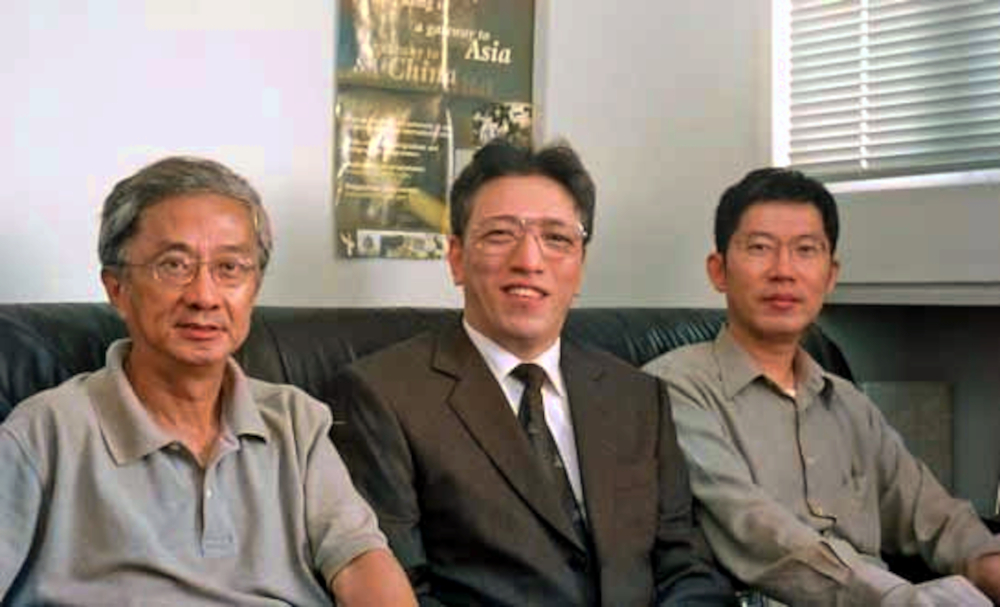}
    \caption{Ning Cai, Robert Li, and Raymond Yeung (from left to right).}
    \label{fig:Cai-Li-Yeung}
\end{figure}

Ning significantly assisted Rudolf Ahlswede teaching his students, drawing on his extensive knowledge and making contributions across various domains. Notably, he collaborated with Christian Deppe on search problems and error-correcting codes. They even shared an office for a period in Bielefeld. Christian vividly recalls moving into the office, where Ning had a bed that he removed upon sharing the space with Christian.

Despite Ahlswede's efforts, Ning's contract in Bielefeld could not be extended after nine years due to concerns from the university administration about potential claims for a permanent position. Nonetheless, Ning remained involved in educating the next generation of Ahlswede's Ph.D. students during his final two years in Bielefeld, mentoring talents like Christian Deppe, Christian Kleinewächter, Lars Bäumer, Andreas Winter, and Gohar Khuregyan.

In 1998, Ning departed Bielefeld and initially assumed a research fellow position at the School of Computing, National University of Singapore. By 2000, he transitioned to a research associate role at the Department of Information Engineering, The Chinese University of Hong Kong, where he continued his collaboration with Raymond Yeung and Bob Li on network coding.
Throughout his journeys, Ning's wife Jinying Zhi and his son Minglai Cai remained in Bielefeld.

Despite finding fulfillment in Hong Kong's research environment, Ning accepted Rudolf Ahlswede's invitation to return to Bielefeld on a research position within Ahlswede's new project at the Center for Interdisciplinary Research (ZIF), University of Bielefeld from 2002 to 2004.

\begin{figure}
    \centering
    \includegraphics[width=10cm]{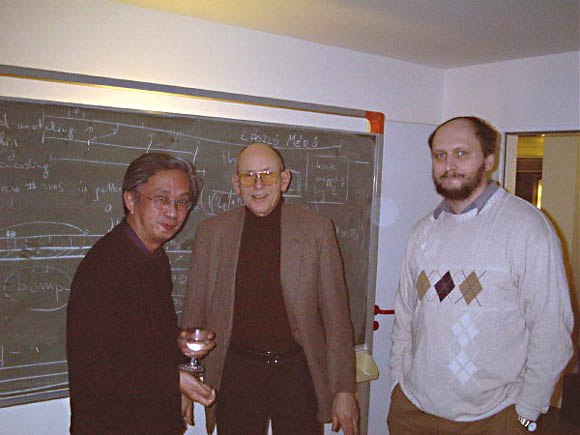}
    \caption{Ning Cai, Edward van der Meulen, Vladimir Blinovsky (from left to right) during a discussion at the ZIF in February 2002.}
    \label{fig:ZIF-2002}
\end{figure}

During that time, Ning began in earnest his contributions to quantum information theory, starting with the most important analysis of the quantum wiretap channel, containing a critique of the use of accessible (Shannon) information and a justification of the use of the Holevo information to define security. This paper, in collaboration with Andreas Winter and Raymond Yeung, gave the now-standard formula for the private capacity of a quantum wiretap channel (in parallel and independently to Devetak), and was published in \emph{Probl. Inf. Transm.} in 2004. Subsequently Ning in collaboration with Rudolf Ahlswede made several significant contributions ti quantum Shannon theory, most importantly their strong converse for classical-quantum multiple access channels, by generalising Ahlswede's wringing technique from the classical case. 

In 2006, after spending another year in Hong Kong, Ning assumed the role of Distinguished Professor at the State Key Laboratory of Integrated Services Network at Xidian University in China.

In April 2016, he relocated to Shanghai, where he held the esteemed position of Distinguished Professor at the School of Information Science and Technology at ShanghaiTech University.

Throughout his tenure in China, Ning Cai held a standing invitation to visit
Holger Boche’s research group at the Technical University in Munich (TUM), where Cai's
son Minglai also pursued doctoral and later postdoctoral positions. He was a visiting professor at TUM every year from 2014 until his death. Typically, Ning arrived in Munich in late autumn and left the TUM for Shanghai in the early spring. During his time in Munich, there were several special events, which the Munich Information Theory groups celebrated together with Ning. On the scientific side, Ning's election as an IEEE Fellow and the aforementioned IEEE Eric E. Sumner Award and ACM SIGMOBILE Test-of-Time Paper Award were properly honoured. Ning was very surprised by the above-mentioned awards and was of course delighted to receive them, as he always focused exclusively on research and has cared about politics in science. The most important personal event for Ning during his stay in Munich was his 70th birthday, which the Munich Information Theory groups also celebrated with him.  During all his stays in Munich, Ning Cai and Holger Boche worked on three fundamental topics of information theory. 

The first topic was the development of a complete theory for arbitrary varying quantum channels. So far, the theory of arbitrary varying quantum channels is a combination of the extension of Ahlswede's theory for arbitrary varying classical channels to quantum channels, and an adaptation of Csiszar and Narayan's theory to quantum channels with classical inputs. This theory of arbitrarily varying quantum channels is sufficient to completely solve the problem of transmitting classical messages via arbitrarily varying quantum channels. For more general quantum communication tasks, however, it is completely open when a positive rate can be achieved for these communication tasks for transmission via arbitrarily varying quantum channels. For this purpose, a theory in the sense of Csisz\'ar and Narayan would have to be developed for arbitrarily varying quantum channels. Overall, this task proved to be very difficult in the sense that no progress could be made despite the time invested.

The second question was the further development of the theory of generalized arbitrarily varying classical and quantum channels. The aim here was to finally develop a rigorous theory for transmission via channels with powerful attackers. This line of research was motivated by the popularity of quantum key distribution protocols and experiments at the time. In this application, robustness against absolutely powerful attackers on the physical communication system was claimed. However, this claim could not be proven in the published results due to the under-formalized modeling of the attacks.

The third question was the achievability of semantic security over generalized arbitrary varying quantum channels. In particular, Ning Cai and Holger Boche were interested in the difference between semantic security and strong security. The motivation for this question again came from the field of quantum key distribution and is closely related to the second question above. Semantic security is preferred as a security measure for quantum key distribution by the regulatory organizations over the security measure of strong security.

\begin{figure}
    \centering
    \includegraphics[width=10cm]{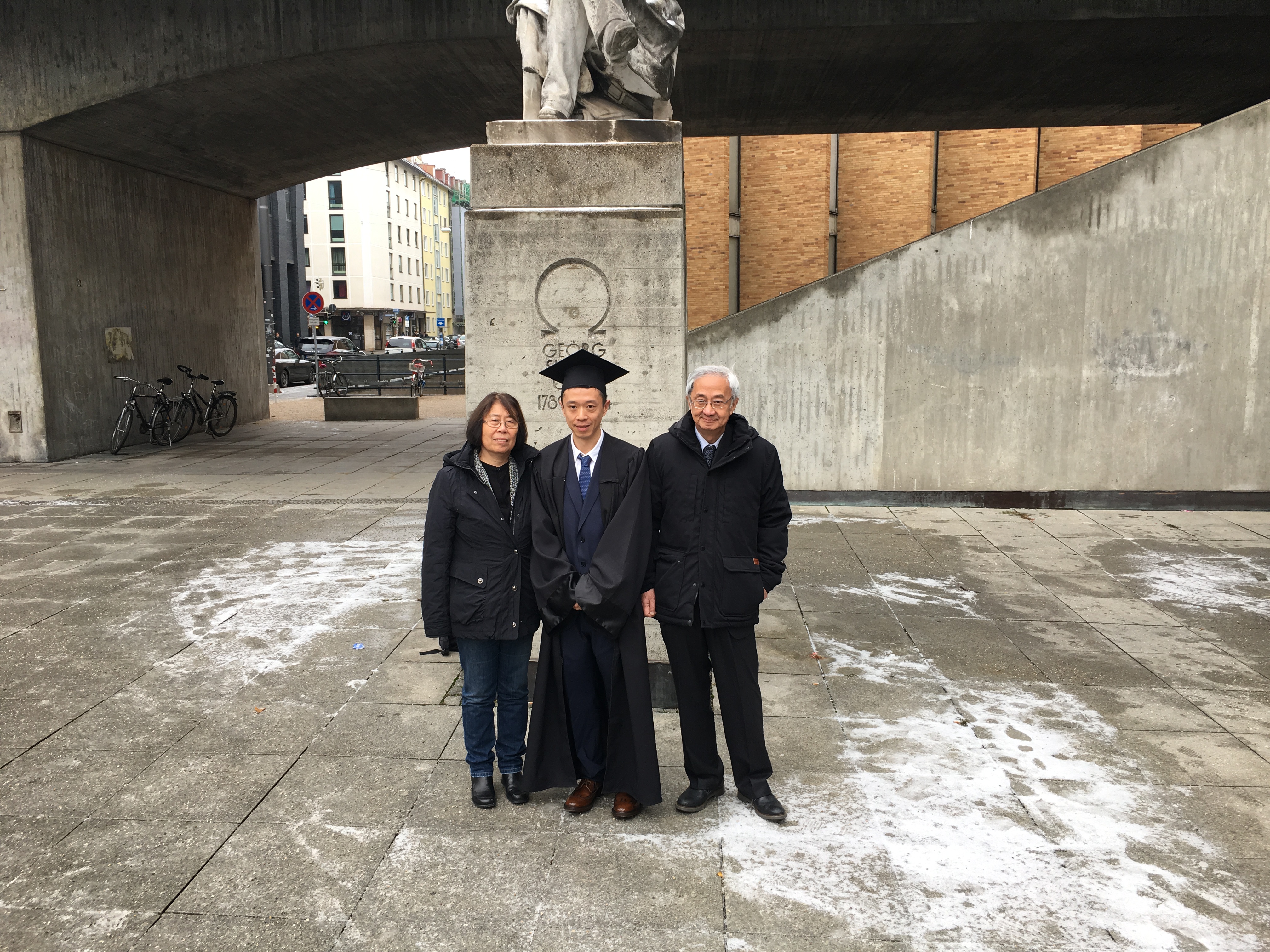}
    \caption{Jinying Zhi (Ning's wife), Minglai Cai (Ning's son), Ning Cai (from left to right) in Munich 2018.}
    \label{fig:Munich-2018}
\end{figure}

Questions two and three above were solved shortly before Ning's death.  The theory corresponding to question 2 could be fully developed, and for question 3, the influence of attacks on the corresponding security measures for generalized arbitrarily varying channels was fully characterized. In particular, the capacity for the semantic security measure can be zero, although the capacity for the strong security measure is much larger than zero. The two scientific papers are close to completion.

Ning Cai was also heavily involved in the other research topics of the Munich Information Theory groups. Ning Cai was very actively involved in Christian Deppe's and Holger Boche's BMBF projects on quantum repeaters, and contributed to many joint publications. He also worked very intensively with his son Minglai Cai. During this time, Minglai Cai worked as a PhD student and later as a post-doc in Holger Boche's group.

Ning Cai was also very interested in fundamental questions of information theory such as multiletter formulae, superactivation and computability of capacity expressions. Ning Cai and the Munich Information Theory groups organized many seminars on the value of multiletter characterisations for capacities. These led to a whole series of publications on continuity of important capacities and also to the first superactivation results in classical information theory.  

In Munich, Ning also showed great interest in the application of information theory to future communication systems. With his questions and comments, he greatly advanced topics such as deterministic identification, multiuser communication and multiuser identification and entanglement-assisted communication in several BMBF projects such as NewCom and 6G-life. Ning was also very active in the information-theoretic foundations for Physical Unclonable Functions and the results were published in several papers in collaboration with Sebastian Baur, Moritz Wiese and Holger Boche.

Ning typically visited Munich during the winter months, allowing him to spend the Christmas holidays with his family in Bielefeld. These visits often included organized meetings with former colleagues from the Ahlswede research group. In 2016, many of these colleagues convened in Barcelona for the ``Beyond IID in Information Theory'' workshop, which Andreas Winter had organized directly after ISIT. Another memorable photograph was captured during this event.

\begin{figure}
    \centering
    \includegraphics[width=10cm]{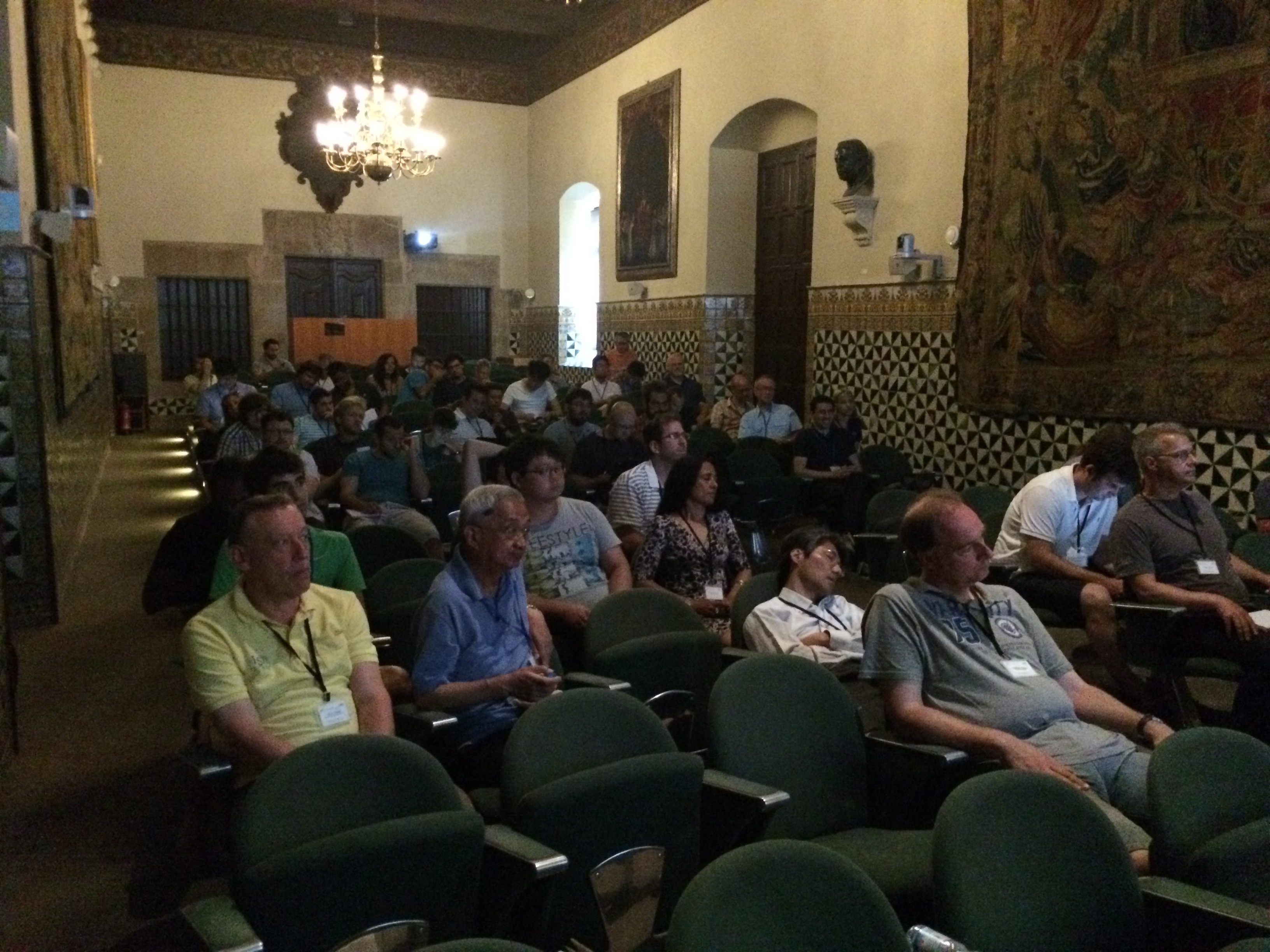}
    \caption{Ning Cai among the audience of Beyond IID in Information Theory, Barcelona 2016.}
    \label{fig:BIID4-2016-Barcelona}
\end{figure}

There was also extraordinary interest in Ning Cai's information-theoretic work from the engineering community not only in Munich but also in wider Germany. This particular interest was evident at the memorial symposium for Ning Cai at ZIF, where a very large number of participants came from the engineering sciences, and where three of the four BMBF 6G Hubs were very prominently represented. Even start-ups, whose business ideas are based on the fundamental contributions of Ning Cai, Bob Li, Raymond Yeung and Rudolf Ahlswede on network coding, took part in the ZIF Memorial Symposium for Ning Cai.

\begin{figure}
    \centering
    \includegraphics[width=10cm]{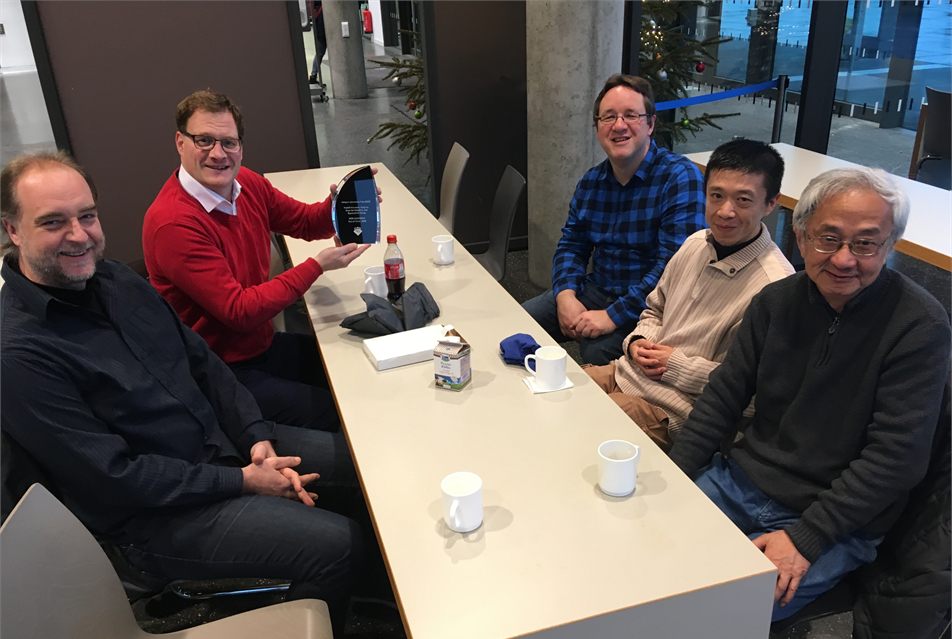}
    \caption{Christian Deppe, Alexander Ahlswede, Christian Kleinewächter, Minglai Cai, Ning Cai (from left to right) in Bielefeld 2019.}
    \label{fig:Bielefeld-2019}
\end{figure}

Amid the rigorous lockdown restrictions in China enforced during the COVID pandemic, Ning found himself unable to conduct teaching duties in Shanghai for three years, prompting him to stay in Bielefeld. Throughout this period, he sporadically met with former colleagues and friends. As per tradition, he secured a table for dinner in January. However, despite typically participating in colleagues' gatherings, Ning declined the invitation this time. When offered the opportunity to reschedule, assuming he was preoccupied, he firmly insisted to Ulrich Tamm:

\begin{quote}
``Dear Ulrich,
Thank you very much for your consideration. I am sorry for that I may not come on Sunday. Please go ahead to arrange the meeting with the other friends.
I shall stay in Bielefeld for a time and we could meet later when you have more free time.
all the best,
Ning''
\end{quote}

This marked our final interaction with Ning. Looking back now, it seems to us  that his absence was due to a significant health issue, possibly indicating a worsening condition. While we were aware of his lung problems for some time, Ning had assured us that he was recovering well after surgery. Sadly, we never had the chance to see him again. Ning will be deeply missed.

\nocite{*}

\printbibliography

\end{document}